\documentclass{sig-alternate-10pt}

\usepackage{url}
\newtheorem{defn}{Definition}
\newtheorem{example}{Example}
\newtheorem{remark}{Remark}

\newcommand{\Esym}{\mathrm{E}}
\newcommand{\E}[1]{\Esym\left[#1\right]}

\begin{document}

\setlength{\pdfpageheight}{\paperheight}
\setlength{\pdfpagewidth}{\paperwidth}

%
\conferenceinfo{SIGCOMM}{2009, Barcelona, Spain}

\title{Evaluating and Optimising Models of Network Growth}

\numberofauthors{4} %
\author{
\alignauthor Richard G. Clegg \\
       \affaddr{Department of Electrical and Electronic Engineering}\\
       \affaddr{University College London}\\
       \affaddr{London, UK}\\
       \email{richard@richardclegg.org}
\alignauthor Raul Landa\\
       \affaddr{Department of Electrical and Electronic Engineering}\\
       \affaddr{University College London}\\
       \affaddr{London, UK}\\
       \email{rlanda@ee.ucl.ac.uk}\\
\alignauthor Uli Harder\\
       \affaddr{Department of Computing}\\
       \affaddr{Imperial College London}\\
       \affaddr{London, UK}\\
       \email{uh@doc.ic.ac.uk}
\and
\alignauthor Miguel Rio\\
       \affaddr{Department of Electrical and Electronic Engineering}\\
       \affaddr{University College London}\\
       \affaddr{London, UK}\\
       \email{m.rio@ee.ucl.ac.uk}
}

\date{30 January 2009}

\maketitle
\begin{abstract}

This paper presents a statistically sound method for measuring 
the accuracy with which  a probabilistic model 
reflects the growth of a network, and a method for 
optimising parameters in such a model. The technique is 
data-driven, and can be used for the modeling and simulation
of any kind of evolving network. 

The overall framework, a Framework for Evolving 
Topology Analysis (FETA), is tested on data sets collected  
from the Internet AS-level topology, social networking websites and  
a co-authorship network.  Statistical models of the growth of these networks
are produced and tested using a likelihood-based method. The models
are then used to generate artificial topologies with the same
statistical properties as the originals.  This work can be used to 
predict future growth patterns for a known network, 
or to generate artificial models of graph topology evolution for 
simulation purposes.  Particular application examples 
include strategic network planning, 
user profiling in social networks or infrastructure deployment 
in managed overlay-based services. 
\end{abstract}




\section{Introduction}
\label{sec:intro}

In recent years there has been much interest in creating simple probabilistic models
which can be used to produce topologies which replicate 
certain statistical properties of a given target network.  Many of these models  
depend on a procedure by which a network is progressively ``grown'' 
from a small ``seed" (with a handful of links) into an artificial 
topology which is as large as required. If the model is successful, the artificial 
network will have similar properties to the original. Thus, these models rely 
on finding a network evolution model that produces networks 
which are structurally similar to the target network.

In much of the previous research in this field the usual 
way of achieving this is to hypothesise an evolution model for the target network, 
grow an artificial network of at least the same size using that model, 
and compare several key graph theoretical statistics with the respective ones from the target 
network. This is usually done multiple times, so that the expected values 
of the statistics can be obtained. If after this process the model is found to be 
unsatisfactory, it is updated accordingly and the whole process repeated. 

Thus, the development of a topology evolution model following the 
methodology detailed above will require the construction of large numbers 
of ``test'' topologies that use tentative evolution models. Since the construction of these
topologies can be computationally cumbersome if the networks in question are large,
the analysis of network evolution models has not been widely adopted 
in practice.

The main contribution of this work is two-fold. Firstly (and most importantly), 
we present a set of statistics that directly measure the likelihood of a 
given probabilistic evolution model giving rise to a given target network -- no 
``test'' topologies need to be constructed. Secondly, 
we present a framework for exploratory testing and optimisation of certain 
(quite general) classes of network growth models.  

The statistics that our technique produces 
are an unambiguous and statistically
rigorous measure of the likelihood of the evolution of the 
target network arising from any particular 
hypothesised probabilistic model.  
These statistics are quick to produce (much more
so than growing a test network of the same size), and could be used
as a fitness function for state-space searches or genetic algorithms
to automatically optimise parametrised classes of models.

We will refer to this statistical framework as FETA 
(Framework for Evolving Topology Analysis). 

The structure of this paper is as follows.  Section \ref{sec:FETA} describes
the FETA framework in detail.  Section \ref{sec:evaluate} shows how the model 
likelihood is derived, and section \ref{sec:fit} describes the fitting procedure
for optimising model parameters.  Section \ref{sec:network} describes the 
model fitting for the five network examples investigated in this paper.  Section 
\ref{sec:reconstruct} shows how the fitted models can be used
to generate artificial topologies that replicate specific statistical measures 
of the corresponding real networks.

\subsection{Motivation}

The problem of creating artificial topologies with the same growth
dynamics as a target network is an important one.  As networks
grow, their statistical measures change and undesirable emergent properties
may occur.  A good statistical model of how a given target
network grows is an important goal which  has applications
in many fields, but especially in the design and optimisation of 
distributed computation and communication systems.
A tier one network provider may wish to be able to model
the future growth of the AS network to predict and potentially
avoid undesirable network properties, or to strategically choose 
its peering agreements.  
The owner of an online social network may wish to be able
to predict, from their position in the network, consumption patterns or 
demographic characteristics, which users are more likely
to accrue ``friends" and hence influence others. This information can be
used for targeted advertising, marketing or capacity planning purposes.
Finally, a provider of overlay-based services (such as Skype or Akamai) 
may need to plan based upon the future evolution of their overlay network.  
As key network statistics change, they may wish to adapt their protocols, or
to modify infrastructure deployment strategies accordingly.

\subsection{Background}
\label{sec:background}

The field of generating graphs (or \emph{networks} or \emph{topologies}, 
the words seem to be used almost interchangeably in the literature) 
using random processes is usually considered to begin with Erd\H{o}s and 
R\'{e}nyi \cite{ER}.  An early study by Price \cite{price}
found that the degree distribution of co-authorship network of scientific papers
obeyed a power law.  Much later it was 
discovered that the Internet Autonomous System 
(AS) graph also follows a power law \cite{powerlaw} and
this finding was also shown to apply to a large number
of other networks, including social networks, hyperlinked document 
networks and networks derived from biological systems.
The well-known Barab\'asi--Albert (BA) model \cite{ba} 
provided a seminal explanation of scaling network topologies 
in terms of a ``preferential
attachment" model where ``rich get richer": the probability
of connecting to a given node is exactly proportional to its degree.
This led to several papers which attempt to explain network evolution
in terms of node degree and related properties such as the
BA \cite{ba}, ASIM \cite{integrated} and AB \cite{ba2} models. 

Bu and Towsley \cite{bu} introduced the
Generalised Linear Preference (GLP) model, which modifies
the preferential attachment model by raising the degree of the node to
a small power.  Zhou and Mondrag\'on \cite{zhou2004} presented the 
Positive-Feedback Preference (PFP) model, which also modifies
preferential attachment by raising node degree to a small power (but
this power also depends on the node degree).  

It has been shown that a model which faithfully reproduces
the node degree distribution may not capture all the important
properties of a graph \cite{Willinger2002}.  To account for this, the 
ORBIS model \cite{orbis}  reproduces the statistics
of subgraphs of small orders to take account of degree-degree
and higher order characteristics.  The ORBIS model is slightly different
to the growth models which the FETA approach uses, as the model uses
rewiring and rescaling rather than a hypothesised growth model.

The typical assessment of topology generation models has focused on
measuring a number of statistics on the real network data.  Such measures have
included the number of
nodes and links, average and maximum node degree, best-fit power-law exponent,
rich-club connectivity, probability of nodes with low degrees (1, 2 and 3),
characteristic path length, average and maximum triangle coefficient,
average and maximum quadrangle coefficient, average $k_{nn}$
and average and maximum betweenness (see \cite{hamedsurvey} for
definitions of these properties and a review of topology generation from 
an Internet perspective).  A candidate artificial model is then tested
by creating an artificial topology using the model
and seeing how well the topology reproduces several
statistics measured on the real data set.
Occasionally, a new network statistic may be added which existing models
do not reproduce and this can be used to justify a different, improved model.
This approach to model testing and refining based on the generation 
of test topologies and the comparison of a set of statistical measures between 
the test topologies and the target network is characterised here 
as the ``basket of statistics" approach.

\begin{figure*}[ht!]
\begin{center}
\includegraphics[width=11cm]{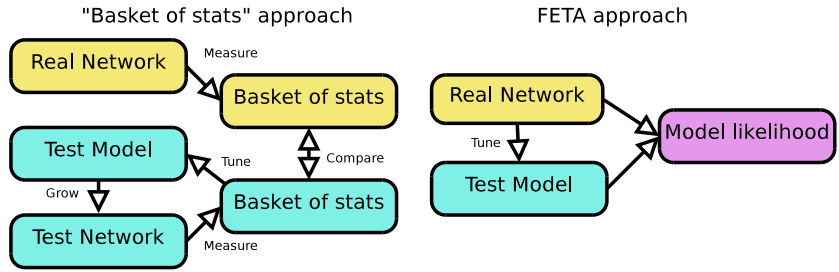}
\end{center}
\caption{The FETA approach compared with the ``basket of statistics" approach.}
\label{fig:FETAapproach}
\end{figure*}

Willinger et al \cite{Willinger2002} called for a ``closing of the loop" between 
the discovery of ``emergent phenomena" and the models which reproduce them.
They emphasise the importance of a ``validation" step to ensure that a particular 
proposed model is consistent with the real data.  The evaluation 
framework given in section \ref{sec:evaluate} 
provides this validation, albeit
at the expense of requiring data about how the 
network evolves rather than a static snapshot.  In so doing, however,
the generation of test topologies is avoided and the processing of bigger graphs
or more complicated models becomes possible.

Figure \ref{fig:FETAapproach} contrasts the approach used by FETA with the
``basket of statistics" approach which has been previously used.  By
directly assessing model likelihood, our approach short circuits the 
cycle of generating and measuring test networks to optimise a test model.

\section{Framework for Evolving Topology Analysis}
\label{sec:FETA}

The Framework for Evolving 
Topology Analysis (FETA) allows the investigation of growth models 
for real networks where 
information is known about the order in which links were added
to the network.  The aim of FETA is to produce probabilistic models
which fit the observed evolution of these networks.  
The class of models which FETA can work with includes BA
\cite{ba}, AB \cite{ba2}, GLP \cite{bu} and PFP \cite{zhou2004}.  

The probabilistic models used by FETA
are described in terms of two components referred to as
an {\em inner model\/} and an {\em outer model\/}.  It is the
inner model which FETA evaluates and fits, which means that
only models based on probabilistically growing 
networks are compatible with FETA.

\begin{defn}
The {\em outer model\/} chooses an operation which 
will make a change to the existing
network.  This could be add and connect a node, add a link
between nodes, delete a node or delete a link between nodes.  The 
{\em inner model\/} defines the probabilities for selecting the 
node or nodes involved in the operation.  
\end{defn}

\begin{defn}
The {\em inner model\/} defines the probabilities for selecting the 
node or nodes involved in the operation defined by the outer model. The 
inner model used by a given evolution model can vary,
depending on whether the outer model operation is a connection to a new node 
or a connection between existing inner
nodes. 
\end{defn}

For example, the AB model would correspond to an outer model which adds 
a new node and then chooses exactly three inner nodes to connect it to, along 
with an inner model that chooses nodes with probabilities proportional to
their node degree.  In PFP and GLP, the outer model is the same as with AB, 
but the inner model is different: the probability is proportional to the degree 
raised to a power.
%

\begin{remark}
As is common in the literature, the main focus of this paper is on
the inner model.  The outer models, where needed in this paper, are
assumed to be of the following simple form: 
a new node is joined to $N$ existing nodes;
following this $M$ (which can be zero)
inner edges are added.  The values of $N$ and $M$
are randomly selected from probability
distributions empirically derived
from the target network.  
\end{remark}

The framework is flexible enough to allow or disallow the possibility 
of multiple edges between the same node pair,  
and to allow or disallow nodes with connections to themselves.  In
this paper only undirected, connected, simple
(no repeated edges and no
edges from a node to itself) graphs are considered.  Removal of nodes
or edges is not considered in this paper.

Section \ref{sec:evaluate} describes the evaluation procedure 
of FETA, that works with any inner model which can assign probabilities
to nodes or edges in a graph.  It will produce statistically reliable 
measures of how well the model fits the observed data.  Section
\ref{sec:fit} describes a fitting procedure which works with 
a subset of inner models which combine sub-models and fit them
using General Linear Models.  Section \ref{sec:inpractice} describes how
FETA is used in practice and gives information about scalability.

\subsection{Model evaluation using FETA}
\label{sec:evaluate}

Consider data from an empirical network which shows how the network grows
in time by the addition of nodes and edges.
This growth data can be decomposed into decisions from the outer model
(whether a link is between existing inner nodes or to a new node) and
the choices of node (which would be controlled by the inner model).

Let $G_0$ 
be the known state of the graph at a certain time.
Assume that the graph is extended by adding edges (sometimes
between existing nodes and sometimes in addition to a new node)
one at a time.
Assume, further, that the state of the graph is known for each one of 
these edge additions up to some step $t$ ($G_0, G_1, \ldots, G_t$ is 
known).
Let $O_i$ be the outer model operation (connect edge to new node or
connect edge between existing nodes) for the $i$th edge addition
since $G_0$. 
Let $I_i$ be the node or nodes selected by the inner model for the
outer model operation $O_i$.  Together $O_i$ and $I_i$ define the
transition between $G_{i-1}$ and $G_i$.  Conversely, if $G_{i-1}$
and $G_i$
are known then $O_i$ and $I_i$ are also known for 
$1 \leq i \leq t$.  The best outer and inner models are those which
best explain $O_i$ and $I_i$, respectively, for the observed periods.
This paper focuses on the selection of the inner model.  

Let
$C$ stand for all of the observed inner model choices 
$I_1, I_2, \ldots, I_t$, and let $\theta$ be some inner model which
attempts to explain the observed inner model choices $C$
in terms of some statistical properties of the graph.  At each step $i$,
$\theta$ maps graph properties (and, perhaps, other properties,
such as whether a new node or an inner edge is being connected, 
or properties associated with the node but exogenous to the 
network topology) to probabilities.

In order to simplify the explanation, 
assume for remainder of this section that the outer model always
involves the choice of a single existing node to connect to a new
node.  In this case $C$ is simply an ordered list of the nodes
chosen at each observed step and $I_i$ is the node
connected at step $i$.  Evaluation of the model $\theta$
is now a matter of calculating the likelihood of $C$ given the
model $\theta$.  The larger this likelihood, the better the model
fits the observed data.

Let $p_i(j | \theta)$ be the probability that inner model
$\theta$ assigns to node $j$ at step $i$.  To be a valid model
$\theta$ should ensure that $\sum_j p_i(j | \theta) = 1$ where
the sum is over nodes.
It follows
that $p_i(I_i | \theta)$ is the likelihood of the choice $I_i$ at
step $i$ given model $\theta$.  The likelihood of all the
observations $C$ given model $\theta$ is given by the product
$$
L(C | \theta)= \prod_{i = 1}^t p_i(I_i | \theta).
$$
It is also useful to define the log likelihood 
$l(C | \theta) = \log(L(C | \theta))$.  The larger
the likelihood (or log likelihood) the better the model
explains $C$.

\begin{defn}
The {\em null model\/} $\theta_0$ is defined as 
the model which gives every
node in the choice set equal probability (this can also be
thought of as the {\em random model\/}).  The {\em saturated
model\/} $\theta_s$ 
is a model with as many parameters as data points.  In this case, 
$\theta_s$ ensures that $p_i(I_i|\theta_s) = 1$ for all 
$i \in \{1, \ldots,t\}$.
Hence $L(C|\theta_s)=1$ and $l(C|\theta_s)= 0$.
\end{defn}

Now it is useful to define some measures of the goodness of
the model using the statistic known as deviance.

\begin{defn}
The {\em deviance\/} of model $\theta$ is minus two 
times the log-likelihood
ratio between the model $\theta$ and the saturated model $\theta_s$,
$$D = -2 (l(C|\theta) - l(C|\theta_s)) = -2l(C|\theta).$$ 
\end{defn}

Evidently, the deviance will always 
be positive (or zero for the saturated model), 
and the smaller it is,  the better the model $\theta$
explains the data.

\begin{defn}
The {\em null deviance\/} $D_0$ of a candidate model $\theta$ 
is given by $$D_0 =  -2 (l(C|\theta) - l(C|\theta_0)).$$
\end{defn}

Thus, $D_0$ will always be negative if the model $\theta$ explains $C$ better
than the null (random) model $\theta_0$.  The smaller $D_0$, the better
$\theta$ explains the choice set $C$.

Because of the size of the data sets used in this work ($|C| \sim 100,000$) 
then the magnitude of $D$ can be quite large
and depends critically on the size of $C$.
It is useful to have a statistic which defined on a more
comprehensible scale, and
invariant to the size of $C$.  We present such new statistic, the 
{\em per choice likelihood ratio\/}:

\begin{defn}
The {\em per choice likelihood ratio\/} $c_0$ is the likelihood ratio between
$\theta$ and the null model $\theta_0$ normalised by the number of choices.
$$c_0= \left[\frac{L(C|\theta)}{L(C|\theta_0)}\right]^{1/t} = 
\exp \left[\frac{l(C|\theta) - l(C|\theta_0)}{t}\right].$$
\end{defn}

The quantity $c_0$ is one if $\theta$ is exactly as good as $\theta_0$, greater
than one if it is better and less than one if it is worse.
Note that $D, D_0$ and $c_0$ are simply different ways of
looking at the model likelihood. 

It should be noted though, that while a lower deviance or a higher
per choice likelihood ratio always indicate a better fit, this alone 
does not mean a model should be preferred.  The saturated model $\theta_s$
gives a perfect fit to data, but it is a useless model for practical 
purposes since it can only reproduce the data it has been given.  What
is needed is a trade off between fit to data and a parsimonious model.
Adding new parameters to a model is only good if the improvement to
the fit (reduction in $D$, increase in $c_0$) justifies the extra parameter.
One criteria would be Akaike's An Information Criterion (AIC) \cite{AIC} 
which
is given by $A = D + 2k$ where $k$ is the number of free parameters in
the model.  However, given the size that $D$ typically attains in this modelling, 
this seems unlikely to prove a useful distinction.

\begin{example}
An example will help comprehension.  Consider an initial
graph which is the two link network
consisting of nodes $\{1,2,3\}$ and edges $\{(1,2),(2,3)\}$.  The
network grows by adding node $4$ and link $(2,4)$ and then node
$5$ and link $(2,5)$.  We assume the simple outer model add one node
and connect it to one existing node at every stage.  The inner model
must explain $C=(2,2)$,  $I_1 = 2$ and $I_2 = 2$ given $G_0$ and $G_1$.
The null
model $\theta_0$ predicts equal probabilities ($1/3$ each)
for node $4$ to connect to nodes $1$, $2$ or $3$ and equal
probabilities of $1/4$ each, for node 5 to connect to nodes 1 to 4.  
Therefore, for this $C$ and the null model
the likelihood is $p_1(I_1|\theta_0) = 1/3$ and
$p_2(I_2|\theta_0) = 1/4$.  The null likelihood $L(C|\theta_0) = 1/12$.
If, on the other hand, we consider $\theta$
as the preferential attachment model (probability of attachment proportional
to node degree) then, given $G_0$ the node probabilities are $(1/4,1/2,1/4)$
and given $G_1$ they are $(1/6,1/2,1/6,1/6)$.
The likelihoods are $p_1(I_1|\theta) = 1/2$ and $p_2(I_2|\theta) = 1/2$
giving a final likelihood $L(C|\theta)= 1/4$.  From this, deviance, 
null deviance and per choice likelihood ratio can be calculated.
Naturally, real data sets will have many more choices and many more nodes.
\end{example}

\begin{remark} \label{rem:inneredges}
The selection of edges from a set of all possible edges would present
a difficult computational problem, as the set of all possible edges increases
approximately as the square of the number of nodes.  This can be avoided
by considering the probability of choosing edge $(n_1, n_2)$ as the probability
of choosing $n_1$ followed by $n_2$, plus the probability of choosing $n_2$ followed
by $n_1$ (assuming $n_1 \neq n_2$).  The second choice set can be narrowed to
avoid self loops and nodes already connected to the first node if a
simple graph is desired.
\end{remark}

\begin{remark} \label{rem:sepmodels}
Separate inner models can be fitted to each type of operation for the outer model.
Therefore, for example, the hypothesis that new nodes connect using preferential
attachment and inner edges connect using PFP can be explored.  The data set $C$ can
be split into two parts, those choices associated with connecting to new nodes and those
choices associated with adding edges between existing nodes.  In this case, the
deviance of the full inner model is the sum of the deviance of the model components, and
the per choice likelihood ratio $c_0$ can be calculated accordingly.  
\end{remark}

\subsection{Model fitting using FETA}
\label{sec:fit}

The deviance and per choice likelihood ratio can determine which inner model is a better fit
for a given data set.  However, for parametrised models, they do not allow the automatic
tuning of parameters.  In this section a method is introduced based upon the statistical
technique of Generalised Linear Models (GLM) which allows certain (linear) parameters
to be automatically tuned for an inner model.  Again, for simplicity 
of discussion, 
this section considers only inner models which connect nodes to a new node.

Consider an inner node model $\theta$.  It may be that the ideal model is
not pure preferential attachment or PFP, but some mixture of these models.  
Further, it follows that probabilities may be affected other factors (both
inherent in the graph topology and exogenous to the topology but available
as a data input).

Let $d_j(i)$ be the degree of node $i$ in graph $G_{j-1}$ (the graph used to make
choice $j$).  Let $p_j(i|\theta)$ be the probability that model $\theta$ assigns
to node $i$ for choice $j$.  For the null model $\theta_0$ then we have that 
$$p_j(i|\theta_0) = C^n_j,$$
where $C^n_j$ is a normalising constant for a given choice (that is, it is
constant for a given $j$) so that the probabilities sum to one over all $i$.
Similarly, for the preferential attachment model, referred to for now as
$\theta_d$, then we have that 
$$p_j(i|\theta_d) = C^d_j d_j(i),$$
where, again, $C^d_j$ is a normalising constant for a given $j$.
Let $T_j(i)$ be the number of triangles ($3$--cycles) node $i$ is part of in graph
$G_{j-1}$.  Now we can consider some hypothetical model $\theta_t$ where connection
probabilities depend upon the triangles,
$$p_j(i|\theta_t) = C^t_j t_j(i),$$
where, again, $C^t_j$ is a constant for fixed $j$.
A model can be considered which is a linear combination of $\theta_0$, $\theta_p$
and $\theta_t$.  Call this hypothesised model $\theta$. For this model we would have 
that 
\begin{equation}\label{eqn:tofit}
p_j(i|\theta) = \beta_0 p_j(i|\theta_0) + \beta_d p_j(i|\theta_d) + 
\beta_t p_j(i|\theta_t),
\end{equation}
where $\beta_0$, $\beta_d$ and $\beta_t$ are all in the range $[0,1]$ and
sum to one.  These constants are the proportion of each of the models 
which contribute to the final model. We use GLM to find the optimal combination of $\beta$
parameters for a given data set. A brief summary of GLM follows.  

Let $y= \{y_1, y_2, \ldots, y_N\}$ be
some set of observed data we desire to model.  Let 
$x^1 = \{x^1_1, x^1_2, \ldots, x^1_N\}$ and $x^2, x^3, \ldots$ (similarly defined)
be sets of observed data that is to be used to explain $y$.  A relationship
is hypothesised which allows $y$ to be estimated in terms of $x^1$, $x^2$ and
so on.  A model is to be fitted of the form
\begin{equation}\label{eqn:glmexample}
y = \beta_0 + \beta_1 x^1 + \beta_2 x^2 + \beta_3 x^3 + \varepsilon,
\end{equation}
where the $\beta_i$ are parameters (not constrained to a range this time)
which give the contribution of the various components to the variable $y$
and $\beta_0$ which is an intercept parameter and $\varepsilon$ is
an error component.
Fitting GLM can be done automatically using a statistical language
such as R\footnote{\url{http://www.r-project.org}}.  Given observed data, this
can be read into R and a GLM fitting procedure can be used to find 
those $\beta$ values which maximise the model likelihood.  In
addition, the fitting procedure produces the model deviance and
estimates of the errors and statistical significance for each of
the model parameters.  If a parameter is not statistically
significant it should usually be removed from the model.

%
%
%
%
%

Let $P_j(i)$ be an indicator variable which is $1$ if and only if 
node $i$ was actually the node picked for choice $j$, and $0$ 
otherwise. The problem of finding the best model in \eqref{eqn:tofit}
becomes the problem of fitting the GLM,
$$
P_j(i) = \beta_0 p_j(i|\theta_0) + \beta_d p_j(i|\theta_d) + 
\beta_t p_j(i|\theta_t) + \varepsilon,
$$
to find the combined model $\theta$ that best predicts the $P_j(i)$. 
Thus, GLM fitting can be used to find the choice of 
$\beta_i$ which maximises the likelihood of this model.  This
is equivalent to finding the $\beta_i$ which gives the
maximum likelihood for $\theta$ since for model $\theta$,
the expectation $\E{P_j(i)} = p_j(i|\theta)$.

This will give the choice of $\beta_i$ which best combine 
the model components into the unified model $\theta$.  If the $\beta_i$
are in the range $[0,1]$ and sum to one, it can be trivially shown
that $\theta$ is a valid probability model as long as $p_j(i|\theta_1)$, 
$p_j(i|\theta_2)$, $\ldots$ are.  

So, for the period between $G_0$ and $G_t$, for each node, a data point is generated with the parameters 
of the graph relevant to the models, and with a $1$ or a $0$
depending on whether that node was the node actually selected 
as an outcome of that choice.  
The procedure has been tested on data from artificially generated networks and
it has been found to be able to successfully recover their $\beta_i$ 
parameters in a wide variety of circumstances.  Certain model 
component combinations might be problematic to fit, however. An example 
of this would be a model constructed from a PFP and a preferential attachment 
component: since these explanatory variables are very similar for most nodes, 
finding a satisfactory mix using GLM is usually extremely hard.

Note that only the $\beta_i$ parameters can be automatically optimised
by the GLM fitting procedure.  Any other parameters such as the $\delta$
in the PFP model must be fitted by other means, such as 
trying a number of parameter choices and
comparing the deviance or per choice likelihood ratio.

\begin{remark}
As pointed out in remark \ref{rem:sepmodels}, separate models can be fitted
to nodes connecting to new nodes and connections to inner edges.  The items
of data are separated by an analysis tool and they are fitted in different
GLM models.
As in remark \ref{rem:inneredges} fitting inner edges causes issues for
the framework.  The choice of inner edges is broken down into the choice
of two nodes.  The choice set for the second node is constrained by removing
from the choice set those nodes which already have a link from the first node.
\end{remark}

\subsection{FETA in practice}
\label{sec:inpractice}

The FETA evaluation process therefore consists of hypothesising inner
models which might fit the evolution of a target network and
calculating their likelihood statistics as shown in section \ref{sec:evaluate}.
The fitting procedure in section \ref{sec:fit} is used as an 
exploratory tool both to tune linear combinations of model components and 
also to provide hints as to which other components might be introduced
(for example, a negative $\beta$ parameter will rarely produce 
a usable model but, for example, if the $\theta_t$ component produced
a negative $\beta_t$ this suggests that the choice mechanism is
avoiding nodes with a high triangle count).

\begin{figure}[ht!]
\includegraphics[width=8cm]{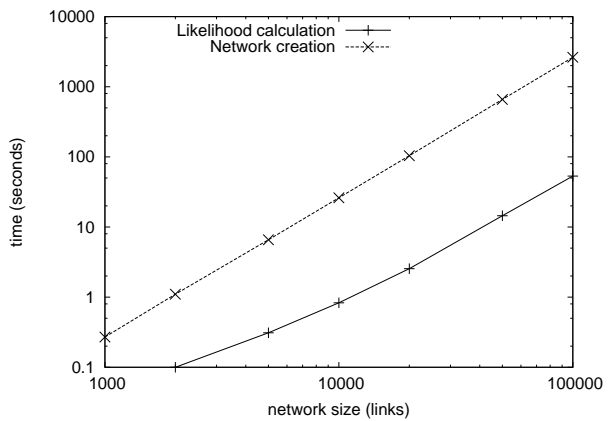} 
\caption{Run time for network creation and analysis processes in FETA.}
\label{fig:runtime}
\end{figure}

The graph in figure
\ref{fig:runtime} shows the run time for measuring model likelihood
(as described in section \ref{sec:evaluate}) and for creating a network file
with a given number of links (using a test model which is part PFP and part
random).  The tests were run on a 2.66GHz quad core Xeon CPU.
The plot is a log-scale showing how run time varies with network size.
For 100,000 links the network creation process takes 2,631 seconds
and for the likelihood estimation process takes 53 seconds.  
Both processes appear to
scale approximately as the square of the number of links (for times
under 1 second the timing information is not accurate).  
Neither process takes a significant amount
of memory.  The relative speed of the evaluation of likelihood statistics
is another benefit of the FETA approach.  To tune the parameters of a
hypothetical parametrised model using the ``basket of statistics"
approach, a new network would have to be grown for every test model.  
This is much more computationally intensive than 
the calculation of likelihood statistics required by FETA.

\section{Fitting models to network data}
\label{sec:network}

The FETA procedure is used to create inner models for several networks of
interest.  Section \ref{sec:pub} fits models to a co-authorship network 
inferred from the arXiv database.  Section \ref{sec:ucla} fits models to a
view of the AS network topology referred to here as the UCLA AS network
and section \ref{sec:routeviews} fits models
to a second view of the AS topology, which we refer to here as the RouteViews
AS network.

Finally, section \ref{sec:gallery} fits network evolution models 
to a network derived from user browsing behaviour, and 
section \ref{sec:flickr} fits models to a social network derived 
from the popular photo sharing site Flickr.

Table \ref{tab:networksummary} 
summarises the networks considered
in terms of total edges, total nodes and the edge/node ratio.

\begin{table}[ht!]
\begin{tabular}{|c|l l l|} \hline 
Network & edges & nodes & edge/node  \\ \hline
arXiv & 15,788 & 9,121 & 1.73 \\
UCLA AS & 93,957 & 29,032 & 3.24\\
RouteViews AS & 94,993 & 33, 804& 2.81 \\
gallery &  50,472  & 26,958  & 1.87 \\
Flickr & 98,931 & 46,557 & 2.13\\ \hline   
\end{tabular}
\caption{Sizes of the networks analysed}
\label{tab:networksummary}
\end{table}

Several model components were considered in a linear combination as described in 
section \ref{sec:fit}.  Those components are listed below, where $p_i$ is the
probability of choosing node $i$ and $k_\alpha$ is a normalising constant such that
$p_i = 1$ when summed over the choice set. Furthermore, $d_i$ is the degree of node $i$ 
and $t_i$ is the triangle count of node $i$.
\begin{itemize}
\item $\theta_0$ -- the null model assumes all nodes have equal probability $p_i = k_n$.
\item $\theta_d$ -- the degree based preferential attachment model assumes node probability $p_i = k_d d_i$.
\item $\theta_t$ -- the triangle count model assumes node probability $p_i = k_t t_i$.
\item $\theta_1$ -- the singleton model assumes node probability $p_i = k_1$ if $d_i = 1$ and 
$p_i = 0$ otherwise.
\item $\theta_2$ -- the doubleton model assumes node probability $p_i = k_2$ if $d_i = 2$ and 
$p_i = 0$ otherwise.
\item $\theta_p^{(\delta)}$ -- the PFP model assumes node probability $p_i = k_p d_i^{1 + \delta \log_{10}(d_i)}$.
\end{itemize}
Note that the PFP model is the only one to require a parameter.

This notation allows the concise description of a linear additive model
in terms of its components.  For example $\theta = 0.1 \theta_1 + 0.9 \theta_p^{(0.04)}$ 
is a model which is has a component from the singleton model (contributing $0.1$ of the
probability) and a component from the PFP model with parameter $\delta=0.04$ 
(contributing $0.9$ of the probability).  In the models $\theta_1$, $\theta_2$
and $\theta_t$ there is a possibility of all nodes being assigned zero
probability (if there are no singletons, doubletons or triangles respectively).
In this case,
$\theta_0$ is substituted for that model component.  This happens on extremely
few occasions and always very early in network construction.  Obviously
a large collection of model components could be tried but a conscious
decision was taken to limit the number of possibilities for this experimentation.

For each data set, three inner models are tried:
\begin{enumerate}
\item a pure preferential attachment
model,
\item a pure PFP model (with an optimally tuned $\delta$ for connections 
to new nodes, and another one for internal edges),
\item the best model found using the techniques from section \ref{sec:fit}.
\end{enumerate}
Model one was picked because the preferential attachment is a reasonable baseline
for improvement.  Model 2 was picked because investigation showed that for
almost every network the PFP model had low deviance.  Model 3 was
picked to show the improvement (if any) possible by using linear combinations
of models.

The outer model was derived simply by calculating empirically from the
network data two distributions.
\begin{enumerate}
\item the number of inner nodes each new node connects to on arrival,
\item the number of inner edges connected between each new node arrival.
\end{enumerate}
These distributions are then used to create the outer model.  This
is simplistic and obviously further research is required to improve the 
techniques to generate this
outer model.

Results are presented using the metrics from section \ref{sec:evaluate}: 
$D$ is the deviance, $D_0$ is the null deviance and $c_0$ is the 
per choice likelihood ratio.  A better model is indicated by lower $D$ and
$D_0$, and by a high $c$.  
The results are broken down into the contribution from the inner model to
connect to new nodes and the inner model for connecting internal edges.

\subsection{Fitting the arXiv data set}
\label{sec:pub}

A publication co-authorship network was obtained from the online academic
publication network arXiv.  The first paper was added in April 1989 and
papers are still being added to this day.
To keep the size manageable, the network was produced just from the  papers 
categorised as math.  The network is a co-authorship network: 
an edge is added when two authors first write a paper together.  In this case,
because it is required that the network remains connected, edges which are
not connected to the largest connected component are 
ignored.  Multiple edges between two authors are not added.  
The processing of this network is far from perfect, only author names 
(rather than unique IDs) are matched.  
Inconsistent naming conventions mean some authors are recorded by first name and 
surname, and some by initial and surname.  To avoid problems matching John Smith, J. Smith
and John W. Smith, the match is on first initial and surname, though it is
clear this will allow some collisions.
One paper\footnote{\url{http://arxiv.org/abs/math/0406190}} was removed from analysis.
The paper has sixty authors, far more than the paper with the next largest number of authors.
Since each author on a paper forms a graph clique with all the other coauthors in that 
same paper, this paper added 1,732 links for which no arrival order significant to the 
evolution of the network could be found. As a size 60 clique would
distort most network statistics, it was rejected as an outlier.



As described in the previous section, three models were tried, a preferential
attachment model, a pure PFP model and the best model found using
the fitting procedure.
Model 2, the best PFP model was, for connections to new nodes, $\theta_p^{(-0.21)}$
and for connections between inner edges  $\theta_p^{(-0.02)}$. 
Model 3, the best model found, was, for new node
connections, $0.881 \theta_p^{(-0.22)} + 0.119 \theta_1$
and for internal node connections the pure PFP model as in model 2
(no better model could be found).

As can be seen, the best model by all measures is model 3.  It is worth
noticing that the inner edge model does not perform significantly
differently between the three (in any case this is the same model for 2 and 3).
With such a small $\delta$ parameter the model is almost the same as
preferential attachment (model 1).  It should also be noticed that the
deviance itself is hard to compare simply because it is such a large number
that the relative differences seem small.  For the preferential attachment model,
(model 1) the new node model was actually worse than the null model and
this can be seen by the fact that $D_0$ is positive and $c_0$ is less than
one.  Overall, the new node model appears to have made few gains relative
to the new node model in all cases ($c_0$ is larger for the inner edge models
than the new node models)
despite the simplicity of the inner edge model.  This suggests that improving
the new node model is the best focus for model improvements in general.
The improvement in new node model $c_0$ from $1.06$ in model 2 to $1.09$ in model
3 seems significant and indicates that the addition of a model reflecting singletons
is useful.  Remembering that singletons in this case are authors with only a single
other co-author, perhaps this is explained by a desire for those authors with a single
co-author to collaborate with other new authors.

\begin{table}[ht!]
\begin{tabular}{|c|c|l l l|} \hline 
Model & component & $D$ & $D_0$ & $c_0$  \\ \hline
Model 1 & New node &  195,000  & -1,510 & 1.06  \\
Model 1 & Inner edge & 118,000 & -4,091 & 1.31 \\
Model 1 & Overall &  312,000 & -5,600  & 1.15  \\ \hline
Model 2 & New node &  194,000 & -2,450  & 1.10   \\
Model 2 & Inner edge &   118,000 &  -4,170   & 1.32   \\
Model 2 & Overall &  311,000 & 6,610  & 1.18 \\ \hline
Model 3 & New node &  193,000 & -3,090  & 1.13  \\
Model 3 & Inner edge &  118,000  & -4,240  & 1.32  \\
Model 3 & Overall & 311,000 & -7,340   & 1.21 \\ \hline 
\end{tabular}
\caption{Three models tested on the arXiv network.}
\label{tab:arxivmodels}
\end{table}

\subsection{Fitting the UCLA AS data set}
\label{sec:ucla}

The data set we refer to here as the UCLA AS data set is a view of
the Internet AS topology 
seen between January 2004 and August 2008.  It comes from the Internet
topology collection\footnote{\url{http://irl.cs.ucla.edu/topology/}}
maintained by Oliviera et. al. \cite{asevolution}. 
These
topologies are updated daily using data sources such as BGP routing
tables and updates from RouteViews,
RIPE,\footnote{\url{http://www.ripe.net/db/irr.html/}}
Abilene\footnote{\url{http://abilene.internet2.edu/}} and LookingGlass
servers. Each node and link is annotated with the times it was first
and last observed during the measurement period.

As previously stated, our network growth model does not include a removal 
process.  On
the other hand, various links and nodes disappear from the UCLA
data set during the time interval under analysis.  To incorporate this into our 
modelling framework, 
the data is preprocessed by removing all edges and nodes
which are not seen in the final sixty days of the data, so that the
final state of the evolution of the network is the AS network as it is in August 2008.
Edges are introduced into the network in the order of their first sighting.
If this would cause the network to become disconnected, their introduction
is delayed until the arrival of other links and nodes allows them to join 
while maintaining a connected network at all times.

\begin{figure}[ht!]
\includegraphics[width=8cm]{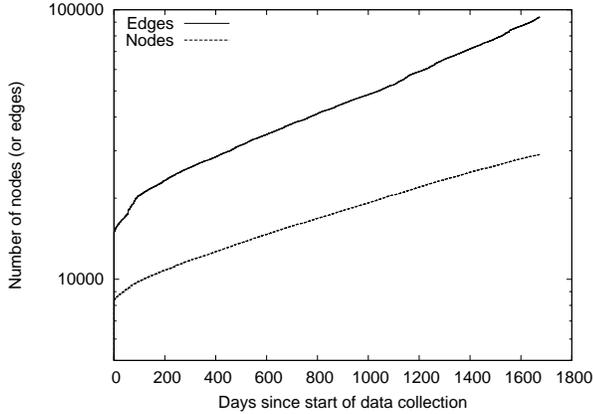} 
\caption{Network growth for UCLA AS network.}
\label{fig:ucla}
\end{figure}

The arrival order of edges 
is only known after timed link arrival data is available in January 2004.  Furthermore,
there is a period of fast discovery of nodes and edges immediately after this 
time where the order of edge arrival is considered to be very uncertain (since 
snapshots are only daily and not every link will be discovered on the first day 
it exists).  Thus, the first days of data are considered a ``warm up'' period and 
removed from the analysis. $G_0$ is taken to be after this warm
up period expires. The 
growth of the network is shown in figure \ref{fig:ucla}.  

Exploratory model fitting on the UCLA AS network showed that a PFP model
was again favoured.  Again
the inner edge model seemed to have a smaller $\delta$ than the new node
model.  No model was found to be a great improvement over PFP but there
was some evidence that including a small singleton component would
make slight improvements.

Three models are tested on the UCLA AS network.  Model 1 is pure
preferential attachment.  Model 2 is pure PFP with different delta parameters
-- $\theta_p^{(0.028)}$ for the new node model and $\theta_p^{(0.007)}$ for the 
inner edge model.
The best model found was only slightly better than this, and it 
and combined PFP with a tiny amount of the singleton model 
The new node model was $0.974\theta_p^{(0.032)}+0.026\theta_1$. 
The inner edge model was 
$0.960 \theta_p^{(0.013)} + 0.040 \theta_1$.

The results of this fitting exercise are shown on table \ref{tab:ucla}.
In this case, the improvement against preferential attachment was
extremely marginal.  It was only model 3 that showed an improvement, 
and this improvement was mostly in the inner edge model (indeed its
new node model was worse than that of model 2).  Overall, the $c_0$ 
values were relatively high indicating a good fit to the data compared
with the random model. 

\begin{table}[ht!]
\begin{tabular}{|c|c|l l l|} \hline 
Model & component    & $D$ & $D_0$ & $c_0$  \\ \hline
Model 1 & New node   & 320,000    & -102,000      &   10.6     \\
Model 1 & Inner edge & 1,790,000   & -402,000      & 5.74       \\
Model 1 & Overall    &  2,110,000   & -504,000      & 6.33        \\ \hline
Model 2 & New node   &  319,000   & -102,000      &  10.8      \\
Model 2 & Inner edge &  1,790,000   & -402,000      & 5.73       \\
Model 2 & Overall    &  2,110,000   & -504,000      & 6.33        \\ \hline
Model 3 & New node   &  320,000   & -102,000      & 10.7       \\
Model 3 & Inner edge &  1,780,000   & -405,000      &  5.82      \\
Model 3 & Overall    &  2,100,000   &  -507,000    &  6.41      \\ \hline
\end{tabular}
\caption{Three models tested on the UCLA AS network.}
\label{tab:ucla}
\end{table}

\subsection{Fitting the RouteViews AS data set}
\label{sec:routeviews}

For the present paper we define the RouteViews AS data set 
as the view of the Internet AS topology 
from the point of view of a single RouteViews data collector.  The raw data 
used to construct it comes from the University of Oregon Route Views 
Project\cite{routeviews}, and it was recovered from 
the parsing of the routing tables obtained by running 
`\texttt{show ip bgp}' on the command line of \emph{route-views3.routeviews.org} 
and capturing the output. To construct the node and link arrival process 
to which we fit our evolution models we process one such table dump 
per day over the time 
interval between April 11th, 2007 and January  16th, 2009. 

It is well known that an AS map obtained in such a way will not be representative 
of the true AS Internet topology (see \cite{as_1, as_2, as_3, as_4}). 
However, a 
validation framework like FETA should be able to discover this difference by fitting 
different growth models to the RouteViews AS data set and the more complete UCLA 
data set.

Since the basic outer models that we set out to evaluate do not have a node 
removal process, we consider only the addition of AS numbers and peerings 
to the AS map, as it is viewed from the perspective of 
\emph{route-views3.routeviews.org}. Thus, we seek to model the cumulative 
AS growth process as viewed from a single BGP peer.

\begin{figure}[ht!]
\includegraphics[width=8cm]{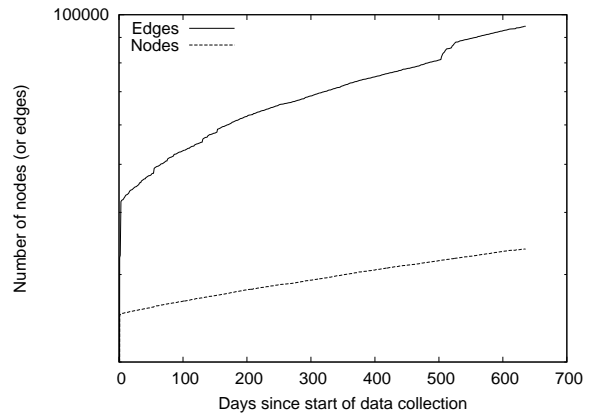} 
\caption{Network growth for RouteViews AS network.}
\label{fig:routeviews}
\end{figure}

As with the UCLA data set, we ignore the very first tables processed, as 
their dynamics are not representative of the system equilibrium growth rate, and their 
timing information is unavailable. The growth of the network is shown in 
figure \ref{fig:routeviews}.  

As before, model fitting on the RouteViews AS data set showed 
that a PFP model was favoured.  As in the previous cases, 
the inner edge model seemed to have a significantly smaller $\delta$ 
than the new node model.  As with the UCLA case, a small singleton 
component in the inner edge model yields slight improvements.

Three models are tested on the RouteViews AS data set.  Model 1 is pure
preferential attachment.  Model 2 is pure PFP with different delta parameters
($\delta=0.034$ for the new node model and $\delta=0.003$ for the 
inner edge model).
Model 3 took the same new node model as model 2, but for the
internal edge model it
combined a pure PFP new node model with the singleton model 
according to $0.87 \theta_p^{(.013)} +  0.13\theta_1$.

The results are shown in table \ref{tab:routeviews}.
As before, the improvement against preferential attachment was
extremely marginal.  It was only model 3 that showed improvement,
as a consequence of a slightly better inner edge model.  Again, $c_0$ 
values were relatively high indicating a good fit to the data compared
with the random model.

\begin{table}[ht!]
\begin{tabular}{|c|c|l l l|} \hline 
Model & component    & $D$ & $D_0$ & $c_0$  \\ \hline
Model 1 & New node   &  138,000   &    -45,400      &    12.7    \\
Model 1 & Inner edge &  1,478,000  &  -257,000    &      4.36  \\
Model 1 & Overall    &      1,620,000 &  -302,400      &      4.81  \\ \hline
Model 2 & New node   &  138,000   &   -46,100    &  13.21      \\
Model 2 & Inner edge &   1,480,000  &   -257,000    &   4.36     \\
Model 2 & Overall    &  1,620,000   &       -303,400 &      4.83  \\ \hline
Model 3 & New node   & 138,000    &  -46,100     &  13.21      \\
Model 3 & Inner edge &  1,470,000   & -264,000   &  4.53         \\
Model 3 & Overall    &  1,610,000   &  -310,100  & 5.00         \\ \hline
\end{tabular}
\caption{Three models tested on the RouteViews AS network.}
\label{tab:routeviews}
\end{table}

As expected, the model found for the RouteViews AS data set is 
quite close to that one found from the UCLA data set, while still being different 
enough to accommodate their differing topological characteristics. Overall, 
the difference in the PFP/singleton mix  
between the best models fitted for the UCLA AS and RouteViews data sets 
suggests that, from the point of view of \emph{route-views3.routeviews.org},
ASs with a single point of attachment to the Internet are more prone to 
becoming multihomed than they are from a more complete perspective of the 
AS topology.


\subsection{Fitting the gallery data set}
\label{sec:gallery}

The website known simply as ``gallery"\footnote{\url{http://gallery.future-i.com/}}
is a photo sharing website. 
To be able to upload pictures and have some control over 
the display of pictures, users have to create an account and login.  
From webserver logs, the path logged in users browse as they move
across the network can be followed. Thus, images become nodes 
in the networks, and a user browsing between two
photos creates a link between the two nodes that represent them. 
These links are overlaid for all users in order to form our network. 

Model 1 is, as usual, a pure preferential attachment model $\theta_d$.
The fitting of model 2 was problematic for this network, minimising 
deviance for the new nodes model with the unusually low delta value of $\delta = 
-1.8$. For inner edges, the PFP model $\theta_p^{(0.015)}$ had lowest deviance.
Finally, Model 3 has the new node model $0.516\theta_d + 0.484\theta_1$, 
and the same inner edge model as model 2 -- that is, $\theta_p^{(0.015)}$.

Table \ref{tab:gallery-user} shows the model likelihood statistics, where 
the inadequacy of the proposed models to the network growth dynamics is 
apparent.  In particular, the
model to connect to new nodes was, for model 1 and model 2, worse
than the null model $\theta_0$ (which connects to nodes at random).
Thus, FETA allows us to discover in a straightforward way that
new node connections in this network do not have a
preferential attachment structure at all.  We hypothesise that the 
peculiar new node arrival process arises from the fact that the
browsing network is, uniquely amongst the networks examined here,
a transient one in the sense that a link between two nodes is made
by a user moving from one picture to the next -- however, no 
permanent record of this is reflected to the user, 
and thus, user behaviour is not influenced by it.

\begin{table}[ht!]
\begin{tabular}{|c|c|l l l|} \hline 
Model & component    & $D$ & $D_0$ & $c_0$  \\ \hline
Model 1 & New node   & 675,000   & 44,300      & 0.523      \\
Model 1 & Inner edge & 586,000   & -17,000     & 1.23       \\
Model 1 & Overall    & 1,260,000    &   27,000    &  0.815      \\ \hline
Model 2 & New node   & 645,000    & 14,000     & 0.810      \\
Model 2 & Inner edge & 586,000    & -17,200      &  1.30      \\
Model 2 & Overall    & 1,230,000   &   -2,750    &  1.02      \\ \hline
Model 3 & New node   & 529,000    & -102,000      &  4.43       \\
Model 3 & Inner edge & 586,000    & -17,000     &  1.30      \\
Model 3 & Overall    & 1,110,000    & -119,000      & 2.43       \\ \hline
\end{tabular}
\caption{Three models tested on the gallery user network.}
\label{tab:gallery-user}
\end{table}

\subsection{Fitting the Flickr data set}
\label{sec:flickr}

The Flickr\footnote{\url{http://flickr.com/}} website allows users
to associate  themselves with other users  by naming them as {\em Contacts\/}. 
In \cite{Mislove08} the authors describe how they collected data for 
the graph made by users as they connect to other users.  The first 100,000 
links of this network is analysed here.  The graph is generated by a
web-crawling spider so the order of arrival of edges is the order
in which the spider moves between the users rather than the order
in which the users made the connections. Thus, the evolution dynamics 
of this network will be determined by the spidering code.

The analysis compares only two different network models, first 
a pure preferential attachment model $\theta_d$; second a PFP model 
with $\theta_p^{(0.05)}$ for the 
new nodes connections
and $\theta_p^{(-0.013)}$ for the inner edge connections.
No combined model was found which improved over the PFP model.

\begin{table}[ht!]
\begin{tabular}{|c|c|l l l|} \hline 
Model & component    & $D$ & $D_0$ & $c_0$  \\ \hline
Model 1 & New node   & 379,000   & -529,000      & 294       \\
Model 1 & Inner edge & 1,600,000    & -479,000      & 9.83      \\
Model 1 & Overall    & 1,970,000    &  -1,010,000     & 27.9       \\ \hline
Model 2 & New node   & 352,000    & -555,000     &  389      \\
Model 2 & Inner edge & 1,590,000    &  -481,000     & 9.93       \\
Model 2 & Overall    & 1,945,733    &  -1,040,000    & 30.7       \\ \hline
\end{tabular}
\caption{Two models tested on the Flickr network.}
\label{tab:flickr}
\end{table}

\subsection{Discussion of model fitting}

Several conclusions can be drawn from the model fitting process. For most 
models considered in this section, providing different model components for the 
inner new node model and the inner edge model yields improved models. Thus,
the separation of the inner model into a sub-model for connections to new 
nodes and a sub-model for new internal edges between existing nodes is usually 
productive.

Moreover, in all but one case (the Flickr data set)
it was found that inner models with a higher likelihood could be obtained
from a linear combination of model components.  Thus, the ability to produce
optimised models through the linear combination of sub-models 
is of use in finding improved network evolution models.

Models based upon PFP generally
had high likelihoods (but this was the only parametrised model
component tried, so this might be simply an issue of 
increased flexibility in the fitting procedure).  The model parameters selected for the two
different AS networks were encouragingly similar, pointing to common network 
evolution dynamics, but had significant differences consistent with the way their 
measurement characteristics.

\section{Artificial Topology Generation}
\label{sec:reconstruct}

The models explored in the the previous section 
have been generated purely by fitting proposed models 
so that their parameters best predicted the actual network
growth process observed. Thus, the models were created 
without growing test networks, measuring statistics on them and 
further refining them -- indeed, the models were identified 
without measuring any statistics about the real network.

However, it is natural to expect that if the null 
deviance $D_0$ and per choice likelihood
ratio $c_0$ predict that model $\theta_A$ is ``better" than 
model $\theta_B$, this will be reflected in model $\theta_A$ 
growing
artificial networks with a better match to the statistics
of the real network than model $\theta_B$. 
Here, therefore, artificial
networks are generated from the seed $G_0$ 
(In the the case of
the AS networks is the state of the network shortly
after measurements started, while in the three remaining 
cases this is simply a single edge).  Each of the models
from the previous section and the random model are
used to grow a network of the same size as the full
real network,  and summary statistics are compared.

The results in this
section need careful interpretation.  In particular, it should
be remembered that the claim is not that these models
are the best possible fit to the real network -- in 
some cases, the claim 
is that the models tried are actually worse than 
simply selecting nodes at random.  The fitting procedure 
in section \ref{sec:fit}  optimises the mixture of model
components (the $\beta$ parameters), while  
the evaluation procedure can optimise other model 
parameters (such as the PFP $\delta$)
using any state space search technique. However, 
the models themselves need to be provided as an input, 
and it may be the case that no perfect model is
to be found from the model components chosen.

However, independently of the precise mathematical 
description of 
the network growth models under test, 
a model with $c_0 > 1$ should be better
than a random model, and the model with the highest $c_0$
should be the best. This is difficult to achieve using 
statistical network measures: saying that one model
reproduces real network statistics ``better" that other 
model is, in itself
problematic.  If a model scores well on three highly
related statistics but extremely badly on two others, is
it a good model?  The ``basket of statistics" does not always
give an unambiguous answer to as to which model is ``best".

For this section, four statistics related to the
degree distribution are used: $d_{\max}$, the
maximum node degree in the network, $d_1$, the proportion of nodes
with degree one, $d_2$, the proportion of nodes of degree
2 and $\overline{d^2}$, the mean square of the node degrees
($\overline{d}$ is a property of the outer model and
automatically equal to that of the real network in all
models here).

In addition, two further statistics are used capturing
the interaction between pairs and triples of
nodes.  The clustering
coefficient $\gamma_i$ of a node is the number of 3-cycles 
that the node belongs to, divided by the potential number of 
3-cycles between its neighbouring nodes (Obviously, 
nodes of degree one do not have any potential triangles and
the clustering coefficient is not defined for them).  In the tables in
this section $\gamma$ is the mean clustering coefficient for
the graph.
The assortativity coefficient $r$ is positive when
nodes attach to nodes of like degree (high degree nodes
attach to each other) and negative when high degree
nodes tend to attach to low degree nodes.
For full definitions of all these quantities see 
\cite{hamedsurvey}.

\subsection{Topology generation using FETA models}

\subsubsection{Statistics on the arXiv evolution model}

\begin{table}[ht!]
\begin{small}
\begin{tabular}{|c|l l l l l l |} \hline 
Model & d=1 & d=2 & $\overline{d^2}$  & $d_{\max}$ & $r$ & $\gamma$ \\ \hline
Real & 0.314 & 0.237 & 31.3  & 127 & 0.00557 & 0.145 \\
Rand & 0.233 & 0.223 & 23.9  & 24 & 0.245 & 0.00285 \\
1 & 0.483 & 0.215 & 123.5  & 446 & -0.060 & 0.0154 \\
2 & 0.431 & 0.204 & 39.1  & 97 & 0.152 & 0.00901 \\
3 & 0.348 & 0.258 & 33.6    & 75 & 0.179 & 0.00748 \\
\hline
\end{tabular}
\end{small}
\caption{Summary statistics arXiv co-authorship network.}
\label{tab:arx_real}
\end{table}

The summary statistics for the arXiv publication network are in table 
\ref{tab:arx_real}.  The previous modelling in table \ref{tab:arxivmodels}
rated model 1 with 
$c_0 = 1.15$, model 2 with $c_0 = 1.18$ and model 3 with $c_0 = 1.21$.
If these figures are reliable, model 3 should be expected to 
be a better fit than model 2 which is in turn a better fit than model 1.
This is certainly bourne out for the degree distribution statistics, 
with model 3 closest for $d=1$, $d=2$ and $\overline{d^2}$ and marginally
worse than model 2 for $d_{\max}$.  All three models generate networks 
which replicate $\gamma$ badly, with model 1 being the closest. With 
respect to assortativity, model 1 is
closest in absolute terms but it
predicts a disassortative network when the actual network is
assortative.  

While these results are not straight-forward
to interpret, the overall picture seems to confirm that model 3 reproduces 
the statistics of the network better than model 2, which in turn 
beats model 1.  The relatively low $c_0$ value means that the models
should not be a dramatic improvement upon the random baseline, and 
this is bourne
out by the statistics (indeed, for model 1 it is arguable whether the
model is even better than random).

\subsubsection{Statistics on the UCLA AS evolution model}

\begin{table}[ht!]
\begin{small}
\begin{tabular}{|c|l l l l l l  |} \hline 
Model & d=1 & d=2 & $\overline{d^2}$  & $d_{\max}$ & $r$ & $\gamma$ \\ \hline
Real & 0.122 & 0.245 & 6,620 & 3,150 & -0.197 & 0.0584 \\
Rand & 0.129 & 0.118 & 210  & 1,199 & -0.00962 & 0.0173 \\
1 & 0.447 & 0.163 & 2,230 & 4,177 & -0.144 & 0.0190 \\
2 & 0.451 & 0.167 & 3,230  & 5,305 & -0.168 & 0.0148 \\
3 & 0.363 & 0.215 & 3,820   & 6,109 & -0.172 & 0.0121 \\
\hline
\end{tabular}
\end{small}
\caption{Summary statistics, UCLA AS network.}
\label{tab:ucla_real}
\end{table}

As detailed in section \ref{sec:ucla}, evolution 
information was not known for the 
early part of the the UCLA AS network growth. 
Therefore, the first 42,000 edges were taken from the original
network, and its evolution followed from this.
The statistics from table \ref{tab:ucla} gave $c_0= 6.33$ for model 1 and model 2, 
but $c_0 = 6.41$ for model 3.  This implies that model 3 should be a 
modest improvement
on models 1 and 2.  This is bourne out by the statistics
in table \ref{tab:ucla_real} for $d=1$, $d=2$ and
$\overline{d^2}$, but for $d_{\max}$ model 3 performs the worst and is
incorrect by some way.  With $r$, model 3 is again the best and relatively close
to the correct value.  Regarding the clustering coefficient $\gamma$, 
model 3 is best but all models are quite far away from the correct value.  
As predicted, model 1 and model 2 are hard to distinguish
using these statistics. Overall, model 3 was best or close to best
in almost all statistics measured,  as the $c_0$ value predicts. 
All models would be expected to be a good improvement on the
random model and this is shown in all statistics except
$d=1$ which random gets nearly exactly.

\subsubsection{Statistics on the RouteViews AS evolution model}

\begin{table}[ht!]
\begin{small}
\begin{tabular}{|c|l l l l l l |} \hline 
Model & d=1 & d=2 & $\overline{d^2}$   & $d_{\max}$ & $r$ & $\gamma$ \\ \hline
Real & 0.203 & 0.363 & 2,110 & 3,294 & -0.186 & 0.00887 \\
Rand &  0.093 & 0.118 & 630  & 2,289 & -0.0710  & 0.00266 \\
1   & 0.342 & 0.185 & 2,130   & 4,172 & -0.154 & 0.00631 \\
2   & 0.350 & 0.187 & 2,520   & 4,637 & -0.165 & 0.00590 \\
3   & 0.118 & 0.358 & 2,610    & 4,844 & -0.163 & 0.00443 \\
\hline
\end{tabular}
\end{small}
\caption{Summary statistics RouteViews AS network.}
\label{tab:route_real}
\end{table}

From table \ref{tab:routeviews}, it would be expected that model 1 ($c_0 = 4.81$)
would be the same as or very 
slightly worse than model 2 ($c_0 = 4.83$)  and model 3 ($c_0 = 5.00$) 
would be slightly better than either.  It is worth noticing that the ratio
of these figures is small and the expected improvement from 1 to 3 is slight.
This hierarchy is bourne out
by the statistics for nodes of degree 1 and degree 2 with model 3 being
considerably better in both cases.   For $\overline{d}$ and $d_{\max}$,
however, the expectation is reversed and for these statistics,
model 1 is better.  Models 2 and 3 are close to each other and the
correct value for $r$ but for $\gamma$ model 1 is better than either.
In the end it is hard to say from these statistics which model is
the best.  The high values of $c_0$ do unambiguously claim that
all models are superior to random by
some way and this is certainly the case.  The random model is the
worst model for all statistics except for $d_{\max}$.

\subsubsection{Statistics on the gallery evolution model}

\begin{table}[ht!]
\begin{small}
\begin{tabular}{|c|l l l l  l l |} \hline 
Model & d=1 & d=2 & $\overline{d^2}$   & $d_{\max}$ & $r$ & $\gamma$ \\ \hline
Real & 0.0132 & 0.473 & 26.3   & 214 & 0.144 & 0.0829 \\
Rand & 0.217 & 0.117 & 210  & 30 & 0.283 & 0.000809 \\
1 & 0.447 & 0.235 & 369  & 1,442 & -0.065 & 0.00689 \\
2 & 0.279 & 0.205 & 38.0  & 277 & 0.160 & 0.00992 \\
3 & 0.0924 & 0.453 & 51.1   & 354 & 0.0708 & 0.00537 \\
\hline
\end{tabular}
\end{small}
\caption{Summary statistics gallery user browsing network.}
\label{tab:gallery_real}
\end{table}

The gallery likelihood table \ref{tab:gallery-user} shows that 
for model 1, $c_0 = 0.815$ (worse than random), for model 2 $c_0 = 1.02$
and for model 3 $c_0 = 2.43$.  This means that, in the
statistics in table \ref{tab:gallery_real}, model 3 should outperform model 2, 
which itself should outperform model 1.  This expectation is largely bourne out by
the degree statistics, with model 1 very inaccurate for all statistics
based on node degree.   However, in this case, it is hard to see the very
clear distinction between model 2 and model 3 which is predicted
by the $c_0$ values.  Model 3 is certainly better at predicting the number
of nodes of degree one and two and does quite well with $\overline{d^2}$
and $d_{\max}$.  However, it remains hard to claim that model 3 
represents the clear improvement in model accuracy that
the $c_0$ statistic would lead us to expect. 

We have a case where model 1 is expected to be worse than
random and model 2 not much better.  This certainly seems 
to match the statistics provided.   The relatively poor
performance of model 3 remains an anomaly of this data set.

\subsubsection{Statistics on the Flickr evolution model}

\begin{table}[ht!]
\begin{small}
\begin{tabular}{|c|l l l l l l |} \hline 
Model & d=1 & d=2 & $\overline{d^2}$   & $d_{\max}$ & $r$ & $\gamma$ \\ \hline
Real & 0.639 & 0.157 & 7,500 & 11,053 & -0.288 & 0.00196 \\
Rand & 0.245 & 0.179 & 32.4  & 35 & 0.341 & 0.000758 \\
1 & 0.560 & 0.172 & 694  & 1,704 & -0.119 & 0.0216 \\
2 & 0.572 & 0.168 & 1,290   & 3,587 & -0.154 & 0.0107 \\
\hline
\end{tabular}
\end{small}
\caption{Summary statistics Flickr spider network.}
\label{tab:flickr_real}
\end{table}

As detailed in section \ref{sec:flickr}, only two models were tried 
for the Flickr data set. Table \ref{tab:flickr}
gives extremely high $c_0$ values for both models, with model 2 being 
slightly better than model 1.  This is definitely reflected in the
statistics in table \ref{tab:flickr_real}
with model 2 being closer to the real data all statistics.  
The models are quite close on many statistics, but fail
to predict the extremely highly connected node with degree 11,053. 
This may be an artifact of the browsing pattern of the spider, which may 
also be reflected in the $\overline{d^2}$ value being incorrect. 
The high $c_0$ values indicate that both models should be considerably
better than the random model, which is certainly true -- the random 
model is extremely bad. These results point towards considerable
structure to the network network evolution which the random model 
fails to capture.

\subsection{Discussion on topology generation}

None of the models tested here were perfect at reproducing the selected
statistics of their respective network data sets.  In the majority of 
cases, the best fitting models reproduced the degree distribution 
related metrics measured here, but finding the maximum degree 
was often difficult.
However, one thing the modelling in this section certainly shows is the
difficulty of distinguishing between models by considering a large number
of, often correlated, statistics. 

Obviously the models tested here could be improved. However, the network
statistics measured did rank the networks in the same order as the statistics
from section \ref{sec:network} (the exception being model 3 in the
gallery data which, while arguably the best model, was not better by the
expected degree).  
This is an important confirmation of the usefulness of the likelihood 
statistic $c_0$ in assessing the fit of network evolution models. The 
gallery data definitely proved an exception to expectations, and this
is perhaps due to the transient nature of this network as discussed in 
section \ref{sec:gallery}.

A general conclusion of this section on the models themselves was that (apart
from the gallery data), as might be expected, the PFP based models outperformed 
the degree based model, and the ``tweaked" models from the fitting process  
in section \ref{sec:fit} (where better models were found) did better still
in most cases.  

For a given network and a given model
the value of $c_0$ did seem (with one exception)
to be an accurate predictor of how well a model would replicate
the statistics of a target network.  However, it is hard to see 
a connection between the magnitude of $c_0$ between networks
and the success in prediction.  For example, the predictions
on the arXiv network seemed very good for model 3 for most
statistics despite the model only having a $c_0$ value of 1.21.

In general, though, relative likelihood statistics for the different models was
reflected in the performance at reproducing representative network statistics.  
Those models with higher likelihoods
(lower deviances) better reproduced the statistics of the target network.
This confirms the usefulness of the framework for automatic model selection.

\section{Conclusions}
\label{sec:conclusions}

In this paper we present FETA, the Framework 
for Evolving Topology Analysis. The most important contribution of 
FETA is a statistically rigorous and unambiguous likelihood
estimate for a model of network evolution that is quick to compute 
and does not require the generation of test networks for its operation.
The method requires a target
network for which the order in which links are added is known
(at least approximately) for a given
period of time.  Given this data, a model $\theta$ which purports to
explain the evolution can be compared either with a second model 
$\theta'$ or with the null (random) model $\theta_0$ as an explanatory 
model
for the link and node arrivals observed in the target network.
The likelihood statistics can be efficiently calculated and could be used,
for example, as a fitness function for a genetic algorithm or 
in state space exploration for parametrised models.

A second contribution is a fitting procedure which allows
the weightings of linear combinations
of models to be tuned automatically to fit the target network.  This
is an exploratory tool and, in addition to providing the weightings
which best combine the models chosen, can guide the user
as to which other model components might appropriate for the
target network.

Five different networks were tested, and several models
were produced for each.  Artificial networks were grown for each model, 
and for each one of these a set of summary statistics were compared against 
measures taken from the real target network.  Models with better likelihood 
estimators were found to have better agreement with the statistics of the
target network.  This confirms that greater accuracy in terms of the 
likelihood estimator corresponds with a 
closer match to the final target network generated.

A great deal of potential future work arises from this paper. 
The outer model (the part of the model which selects whether to
add a node or an internal edge) was not investigated in any depth.
It would be useful to consider the validation and tuning of 
more sophisticated outer models, and which also allowed node 
and edge deletion.

A model form which has more promise than the linear combination
of model components proposed in section \ref{sec:fit} would be 
one with multiplicatively combined model components
(that is a model of the form 
$\theta = \theta_1^{\beta_1} \theta_2^{\beta_2}\cdots$).
Logistic regression would seem a promising framework for this, 
but nontrivial problems exist with normalisation.  
The evaluation framework from section \ref{sec:evaluate}
would, however, work unchanged with this type of model.

In short, the FETA framework is promising for development in
many ways.  The evaluation framework fits a broad class of
models of network evolution and could be a very useful 
tool for researchers wishing to test hypotheses.  The tools
and data used in this paper are freely available for download
and researchers are encouraged to try 
them\footnote{\url{http://www.richardclegg.org/software/FETA}}.

\bibliographystyle{abbrv}
\bibliography{top_sigcomm2009}
\end{document}